\documentclass[10pt,a4paper,conference]{ieeeconf}
\usepackage{amsmath,amsfonts,stmaryrd,blackboard}
\IEEEoverridecommandlockouts
\def\vec#1{{\bf#1}}
\def\op#1{\hat{#1}}
\def\ket#1{| #1 \rangle}
\def\bra#1{\langle #1 |}
\def\ip#1#2{\langle #1 \mid #2 \rangle}

\def\comm#1#2{\left[ #1, #2 \right]}
\def\ave#1{\langle #1 \rangle}
\def\norm#1{\| #1 \|}

\def\Tr{\operatorname{Tr}}

\def\diag{\operatorname{diag}}

\def\SU{\mathfrak{SU}}

\font\liouvop=eufm10
\def\H{\mathcal{H}}
\def\L{{\text{\liouvop L}}}

\def\RR{\mathbb{R}}
\def\NN{\mathbb{N}}
\def\CC{\mathbb{C}}

\def\J{\mathcal{J}}
\def\eps{\epsilon}

\newcounter{exNo}

\usepackage[pdftex]{graphicx}
\newif\ifpdflatex\pdflatextrue
\makeatletter \@ifundefined{pdfoutput}{\pdflatexfalse} \makeatother
\def\myincludegraphics[#1]#2#3{%
    \ifpdflatex \includegraphics[#1]{#2}
    \else       \includegraphics[#1]{#3}
    \fi}

\title{Comparative Analysis of Control Strategies}
\author{Sonia G.~Schirmer, Peter J.~Pemberton-Ross, Xiaoting Wang %
\thanks{SGS is currently an EPSRC Advanced Research Fellow and also
        acknowledges indirect support from the EPSRC-funded UK QIP-IRC and
	Hitachi.  PJP acknowledges support from an EPSRC Project PhD 
        Studentship and XW thanks the Cambridge Trust for an Overseas 
        PhD Studenthip.}
\thanks{The authors are affiliated with the
        Dept of Applied Maths and Theoretical Physics, 
        University of Cambridge, Wilberforce Road, Cambridge, CB3 0WA, UK.
	Correspondence should be addressed to \texttt{sgs29@cam.ac.uk}.} }

\begin{document}
\thispagestyle{empty}
\maketitle

\begin{abstract}
Different ways of modelling quantum control systems, formulating control
problems and solving the resulting problems are considered and compared.
In particular, we compare the performance of geometric and optimal control,
as well as iterative techniques for optimal control design versus local
gradient optimization using a Lyapunov-type potential function for two 
problems of general interest: global control of qubits and entanglement 
generation in the form of Bell state preparation.
\end{abstract}

\section{Introduction}

Motivated in part by the rapid growth of nanofabrication techniques and
nanotechnology, as well as a surge of interest in novel applications of
quantum effects such as quantum information processing, the control of 
quantum systems is a subject that has received considerable attention 
recently.  Many control strategies for quantum systems have been proposed,
from open-loop strategies for Hamiltonian engineering involving usually
off-line design of control fields based on a model of the system, assuming 
knowledge of the initial state and the control objective, to closed-loop 
strategies mostly based on conditional quantum trajectories and continuous
weak measurements~\cite{PRA49p2133}. Some variants of open-loop Hamiltonian 
engineering have enjoyed considerable success in various experimental 
settings~\cite{NAT417p533,SCI282p919} in the form of adaptive open-loop 
techniques such as direct laboratory optimization~\cite{PRE70n016704,
PRA63n063412}. However, despite the obvious practical importance of 
selecting the best strategy, very little work has been done in comparing 
different control strategies in terms of their effectiveness, efficiency 
and robustness.  This is the topic of this paper.  We will restrict 
ourselves here to comparing open-loop control strategies, and focus 
specifically on geometric, Lyapunov and iterative optimal control design 
techniques.

There are many ways of formulating problems in quantum control as 
optimal control problems, from time-optimal control~\cite{PRA63n032308} 
to variational techniques~\cite{JCP110p9825,PRA61n012101,JCP118p8191,
JMR172p296} incorporating various constraints, e.g., on pulse energies, 
amplitudes and frequencies.  In addition, many quantum control problems 
can be formulated either as state control problems (using either a pure
state vector, wave-function, density operator or real coherence vector 
representation), as optimization of observables problems, or as process 
control problems (e.g.~producing a desired quantum logic gate).  
Unfortunately, although these mathematical models are often equally valid, 
if not equivalent, ways of describing the system and control objective, 
we find that different formulations of the problem can lead to different 
solutions for the optimal control field.  This is significant, as the 
solutions are not always equally desirable from a practical or physical 
point of view.  Different formulations have different complexities when 
it comes to solving the resulting equations, either numerically or 
analytically. 

Finally, once a particular model and problem formulation have been chosen, 
there are various ways of solving the resulting control problem, ranging 
from direct single-step control design methods based on Lyapunov potential 
functions~\cite{Mirrahimi04,qph0601016}, to iterative techniques for 
control field design using gradient-based optimization algorithms such as 
GRAPE~\cite{JMR172p296} and other monotonically convergent algorithms%
~\cite{JCP110p9825,PRA61n012101, JCP118p8191}.  Unfortunately, again, even 
if the formulation of the control 
problem is the same, different techniques can yield very different 
solutions.  The choice of algorithm for a particular control problem may 
depend on further factors.  Certain algorithms are more flexible, allowing 
more realistic models to be used, such as the inclusion of non-Hamiltonian 
processes or a realistic restriction on the capabilities of the pulse-shaping 
equipment.  Other algorithms are numerically and computationally far simpler%
---finding a global optimal solution is simpler for geometric control (where 
it is equivalent to finding a geodesic, and has an analytic solution for many 
systems) than for an optimal control problem (where the numerical solution 
and discretization of the solution is non-trivial to globally optimize).  
The techniques also have different degrees of robustness to errors in our 
models of the systems.  These competing issues of numerical accuracy, facility 
of implementation and robustness to modelling and systematic error must be 
appreciated to find the best practical solution. 

Since we obviously cannot address all of these problems in a single paper, 
we will illustrate some of the issues considering two typical problems of 
significance in quantum computing scenarios: simultaneous, selective control 
of several qubits using global control pulses, and controlled entanglement 
of coupled qubits.  We will compare geometric and optimal control design, 
based on direct optimization of a Lyapunov function and iterative optimal 
control design using monotonically convergent algorithms.  Before we delve
into comparing different control design techniques in Sections~\ref{sec:geom:opt}
and \ref{sec:opt:lya}, we briefly consider different ways of modelling 
quantum control systems and formulating quantum control problems in 
Sec~\ref{sec:models}, and review the relevant control design techniques 
in Sec.~\ref{sec:techniques}.

\section{Modelling quantum control systems \& control problems}
\label{sec:models}

As eluded to in the introduction, quantum systems can be modelled in 
various ways.  Systems initially in a pure state, i.e., unentangled with
their environment and governed by Hamiltonian evolution can be described 
by a wavefunction encoding information about the state of the system that
evolves according to a control-dependent Schrodinger equation
\begin{equation} \label{eq:SE}
   i\hbar \ket{\Psi(\vec{x},t)} = H[\vec{x},\vec{f}(t)] \ket{\Psi(\vec{x},t)}
\end{equation}
where $H[\vec{x},\vec{f}(t)]$ is the control-dependent Hamiltonian and 
$\vec{f}$ the control field applied.  For convenience we will choose units 
here so that $\hbar=1$.  In general, the wavefunction of the system is a 
normalizable and differentiable complex-valued function of time $t$ and 
space $\vec{x}$, and $\op{H}$ is a partial differential operator, but 
Eq.~(\ref{eq:SE}) can be recast as an ordinary differential equation, e.g., 
by expanding $\ket{\Psi(\vec{x},t)}$ in terms of the eigenstates $\ket{n}$ 
of $\op{H}_0$, the system's intrinsic Hamiltonian, $\ket{\Psi(\vec{x},t)} 
= \sum_{n=1}^N c_n \ket{n}$, where $N$ is the dimension of the Hilbert space 
$\H$, in which case we simply obtain a linear matrix ODE
\begin{equation} \label{eq:SE2}
  \dot{\vec{c}}(t) = -i \op{H}[\vec{f}(t)] \vec{c}(t),
\end{equation}
where $\vec{c}$ is a complex column vector, usually normalized to unity, 
$\norm{\vec{c}}=1$, and $\op{H}[\vec{f}(t)]$ is simply a control-dependent
$N \times N$ Hermitian matrix.  The dependence of the Hamiltonian on the 
control fields is furthermore often assumed to be linear so that
\begin{equation} \label{eq:Hlin}
  \op{H}[\vec{f}(t)] = \op{H}_0 + \sum_{m=1}^M f_m(t) \op{H}_m, \quad M<\infty.
\end{equation}
This the standard model for pure-state quantum control.  

We can also represent the state of the system by an operator $\op{\rho}(t)$ 
acting on the Hilbert space $\H$.  For a pure-state system represented by 
$\ket{\Psi(\vec{x},t)} \equiv \vec{c}(t)$, we have simply $\op{\rho}(t) = 
\ket{\Psi(\vec{x},t)}\bra{\Psi(\vec{x},t)}\equiv \vec{c}\vec{c}^\dagger$, 
where $\vec{c}^\dagger$ is the Hermitian conjugate of $\vec{c}$, from which 
we can immediately deduce the evolution equation
\begin{equation} \label{eq:LE}
  \dot{\op{\rho}}(t) = -i \left[\op{H}[\vec{f}(t)],\op{\rho}(t)\right],
\end{equation}
where $[A,B]=AB-BA$ is the matrix commutator.  For a pure-state system 
subject to Hamiltonian evolution the latter formulation is equivalent to 
Eq.~(\ref{eq:SE2}).  Its main advantage is that it can be extended to 
systems initially entangled with their environment, i.e., in a non-pure 
quantum state, or can interact (non-coherently) with their environment 
by adding a non-Hermitian super-operator $\L_D[\op{\rho}(t)]$ to the RHS 
of Eq.~(\ref{eq:LE}).  

Finally, we can represent the state of a quantum system in terms of the
(real) expectation values of a complete set of (Hermitian) observables.
There are obviously many choices for the set of observables but a very
convenient choice are the normalized Pauli matrices
\begin{eqnarray*}
    \op{\sigma}_{rs}^x  &=& \frac{1}{\sqrt{2}}( \ket{r}\bra{s} + \ket{s}\bra{r})\\
    \op{\sigma}_{rs}^y  &=& \frac{i}{\sqrt{2}}(-\ket{r}\bra{s} + \ket{s}\bra{r})\\
    \op{\sigma}_{r}^z   &=& \sqrt{\frac{1}{r+r^2}} 
    \left(\sum_{k=1}^r \ket{k}\bra{k} - r\ket{r+1}\bra{r+1} \right)
\end{eqnarray*}
for $1\le r\le N-1$ and $r <s \le N$.  It is easy to show that the $N^2-1$ 
trace-zero matrices $\op{\sigma}_k$ above, together with the normalized
identity operator $\op{\sigma}_0=\frac{1}{\sqrt{N}}\op{I}$, form a complete 
orthonormal basis for the Hermitian operators on $\H$, and hence we have 
$\op{\rho}=\sum_{n=0}^{N^2-1} s_n \op{\sigma}_k$ with $s_k =\ip{\op{\sigma}_k}
{\op{\rho}}=\Tr(\op{\sigma}_k^\dagger\op{\rho})$.  Thus, we can represent any 
quantum state $\op{\rho}$ by a real $N^2-1$ vector $\vec{s}=(s_k)_{k=1}^{N^2-1}$, 
where $s_0=1/\sqrt{N}$ is omitted as it is constant.  The vector $\vec{s}$ is 
generally called the Bloch or coherence vector in the physical literature, or
the Stokes tensor in the mathematical literature.  This Bloch representation is 
very popular especially for two-level systems as it permits visualization of 
quantum states as points inside a ball in $\RR^3$.~\footnote{In the case 
$N=2$, it is customary to normalize the Pauli matrices such that $\op{\sigma}_k
\op{\sigma}_\ell=2\delta_{k\ell}$ in order to obtain the unit ball in $\RR^3$,
but this is a minor detail.}  It is very easy to see that Eq.~(\ref{eq:LE}) 
thus gives rise to the so-called Bloch equation
\begin{equation}
  \dot{\vec{s}}(t) = A[\vec{f}(t)] \vec{s}(t),
\end{equation}
and Eq.~(\ref{eq:Hlin}) implies $A=A_0+\sum_{m=1}^M f(t) A_m$, where $A_m$ 
are real-orthogonal matrices given by $A_m=\mbox{ad}_{\op{H}_m}$ for $m=0,
\ldots,M$, for Hamiltonian systems.  For dissipative systems one can derive 
an affine-linear Bloch equation $\dot{\vec{s}}(t)=A[\vec{f}(t)]\vec{s}(t) + 
\vec{b}$, although we shall not discuss this case here. 

As this brief overview shows, there are several ways of modelling quantum 
systems, which are often equally valid or even mathematically equivalent 
descriptions of the system.  We shall see, however, that they are not 
necessarily equally desirable from a practical point of view, and depending 
on the problem considered, some may in fact be highly preferable to others.  

But even if we have chosen a particular way to model the system, most quantum 
control problems can be formulated in many different ways.  For instance, the 
three `canonical' control problems
\begin{itemize}
\item \emph{State control:} Steer the system from a given initial state 
      to a desired target state
\item \emph{Process control:} Implement a desired quantum process 
      (unitary operator for Hamiltonian systems)
\item \emph{Observable control:} Drive the system such as to optimize the
      expectation value of an observable
\end{itemize}
are inter-convertible.  E.g., a pure state control problem can be formulated 
as observable optimization problem where the observable is the projector onto
the desired pure state.  A mixed-state control problem can be formulated as
an observable optimization problem with $\op{A}=\op{\rho}$.  An observable 
optimization problem can be translated into the problem of simultaneously 
driving a set of basis states to coincide with the desired eigenstates of the
observable, etc.  In addition to many obvious choices, there are often less 
obvious ones that may turn out to be easier to solve than the more obvious 
choices.


\section{Control design techniques for quantum systems}
\label{sec:techniques}

Before we compare control design techniques, we briefly review the relevant 
techniques.
The simplest control design strategy, with a long history of successful 
application in nuclear magnetic resonance experiments, is the use of 
geometric control pulse sequences.  Although it is rarely formulated this 
way in NMR spectroscopy, the basic idea of geometric control is to find 
a target (unitary) operator that will achieve the control objective---be 
that the implementation of a particular process, or the preparation of a 
quantum state, etc---and to decompose or factorize it, using standard 
techniques such as the generalized Euler angle or Cartan decomposition, into 
elementary operations that can be practically realized by applying simple 
geometric control pulses such as piecewise constant controls or Gaussian 
pulses, for example.  Typically, geometric control sequences can be 
visualized as performing a sequence of rotations on Bloch vectors in 
$\RR^{N^2-1}$, whence the name geometric control.  

An alternative to simple geometric pulse sequences is the use of temporally
or spectrally shaped pulses optimized for a particular task.  The design
of optimally shaped pulses is a non-trivial problem, which is usually solved 
numerically using one of a number of (similar) iterative techniques~\cite{JCP110p9825,
PRA61n012101,JCP118p8191,JMR172p296}.  The particular algorithm we will 
employ in the following involves iteratively solving an \emph{initial 
value problem} for a variational trial function $S_v^{(n)}(t)$ representing 
the state of the system
\begin{equation} \label{eq:rhon} 
  \dot{S}_v^{(n)}(t) = \L[\vec{f}^{(n)}(t),S_v^{(n)}(t)],  
   \qquad S_v^{(n)}(t_0) = S_0,
\end{equation}
followed by a \emph{final value problem} for a variational trial function 
$A_v^{(n)}(t)$ (usually representing an observable)
\begin{equation} \label{eq:An}
  \dot{A}_v^{(n)}(t) = \L[\vec{f}^{n}(t), A_v^{(n)}(t)],  \qquad
  A_v^{(n)}(t_F) = A_F,
\end{equation}
while updating the control field in each step according to
\begin{eqnarray*} 
  f^{(n)}_m(t) &=& (1-\alpha)\tilde{f}_m^{(n-1)} -\frac{i\alpha}{\lambda} 
          \frac{\delta \J(A_v^{(n-1)},S_v^{(n)},f_m^{(n)})}{\delta f_m}  \\
  \tilde{f}_m^{(n)}(t) &=& (1-\beta)f_m^{(n)} -\frac{i\beta}{\lambda} 
          \frac{\delta \J(A_v^{(n)},S_v^{(n)},f_m^{(n)})}{\delta f_m}  
\end{eqnarray*}
starting with an initial trial field $\vec{f}^{(0)}(t)$, and setting
\begin{equation}
   \frac{\delta \J(A_v,S_v,f_m)}{\delta f_m}  
   \equiv \ip{A_v(t)}{\partial_{f_m(t)} \L[\vec{f}(t),S_v(t)]}.
\end{equation}

In the abstract notation above, $S_v$ could be a pure state $\ket{\Psi_v}$, 
a density matrix $\op{\rho}_v$, a real Bloch vector $\vec{s}_v$, or even a 
unitary operator $\op{U}_v$; $A_v$ is a suitable conjugate variable.  The 
super-operator $\L[\vec{f}(t),S_v(t)]$ depends on the chosen model.  For a 
Hamiltonian pure-state system
\[ 
  \L[\vec{f}(t),S_v(t)] \equiv -i\op{H}[\vec{f}(t)]\ket{\Psi_v(t)},
\] 
for a general mixed-state system 
\[
 \L[\vec{f}(t),S_v(t)] 
  \equiv -i\comm{\op{H}[\vec{f}(t)]}{\op{\rho}_v(t)}+\L_D[\op{\rho}_v(t)], 
\]
and for a unitary operator control problem $-i\op{H}[\vec{f}(t)]\op{U}_v(t)$, 
etc.  For a system with control-linear Hamiltonian~(\ref{eq:Hlin}) the 
derivatives of the dynamic operator with respect to $f_m(t)$ are explicitly
\begin{equation}
 \ip{A_v(t)}{\partial_{f_m}\L[\vec{f}(t),S_v(t)]}
 \equiv \bra{\chi_v(t)} \op{H}_m \ket{\Psi_v(t)}
\end{equation}
for a pure-state system, 
\begin{equation}
 \ip{A_v(t)}{\partial_{f_m}\L[\vec{f}(t),S_v(t)]}
 \equiv \Tr\left(A_v(t) [\op{H}_m,\rho_v(t)] \right),
\end{equation}
for a mixed-state system, etc.  $\alpha$, $\beta$, $\lambda$ and the target 
time $t_F$ are non-negative real parameters with $0\le \alpha,\beta <2$ to 
ensure convergence.  It is possible to let $\alpha$, $\beta$ and $\lambda$
be positive functions rather than constants but we shall assume constant
values in this paper.

The third technique we consider is model-based (open-loop) control design 
using Lyapunov functions.  This approach essentially involves choosing 
the control field at every point in time such as to ensure the dynamical 
evolution results in the monotonic decrease of a Lyapunov potential $V$, 
which has a minimum at our target state.  If the objective is to reach a 
given target state, an obvious choice for $V$ is a monotonic function of 
the distance between the current and target states of the system, e.g., 
$V(\rho,\rho_d)=\norm{\rho-\rho_d}^2$, if we use the density operator 
formulation, $V(\psi,\psi_d)=\norm{\psi-\psi_d}^2$ in the pure state 
formulation, or $V(\vec{s},\vec{s}_d)=\norm{\vec{s}-\vec{s}_d}^2$ if 
we use the Bloch vector representation of the state.  Regardless of the 
precise formulation, one easily obtains a simple rule for choosing the 
control field.  For instance, in the Bloch representation, given a system 
satisfying $\dot{\vec{s}}(t)=A[f(t)]\vec{s}(t)$ with $A[f(t)]=A_0+f(t)A_1$ 
where $A_0$ and $A_1$ are dynamical generators, we can easily obtain the 
rule $f(t)=\kappa\vec{s}_d(t)^T A_1 \vec{s}(t)$, where $\kappa$ is a 
positive constant.  The obvious advantage of this approach is that we 
obtain an explicit control law, and under certain conditions, asymptotic 
convergence from a near-global set of target states to the target state 
can be proved~\cite{Mirrahimi04,qph0601016}.

\section{Geometric vs optimal control}
\label{sec:geom:opt}

\begin{figure}
\includegraphics[width=0.5\textwidth]{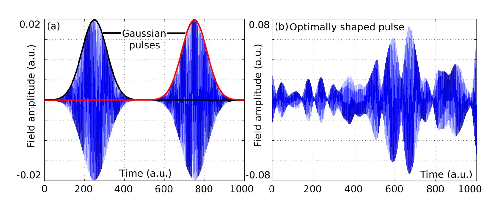}
\vspace{-2ex}
\caption{(a) Two Gaussian wave-packets with `pulse area $\pi$' with frequencies
$\omega_1$ (pulse 1, black envelope) and $\omega_3$ (pulse 2, red envelope), 
respectively. (b) Optimally shaped pulse to achieve simultaneous selective 
excitation of dots $1$ and $3$ while minimizing the (final) excitation of all 
other dots.} \vspace{-2ex}\label{fig:QD13Fields} 
\end{figure}

As an explicit problem in this section we consider an ensemble of five 
quantum dots, which will be modelled as a two-level systems (qubits) with
slightly different energy levels.  Obviously, this is a simplification.
Real systems will have more energy levels but it is a good model system
to study the limitations of frequency selective geometric control and
explore what improvements might be gained using optimally shaped pulses.
The control objective is simultaneous, selective control of all qubits 
using global control pulses, i.e., the ability to simultaneously perform 
selective gate operations on several qubits without any
local addressing.  For the first example we consider local operations 
and assume negligible inter-qubit coupling on the time-scales relevant 
for single qubit gates.

If the individual qubits have different resonance frequencies, due to
variations in size, shape or composition of the dots (as fabricated 
systems such as quantum dots usually do), the simplest control approach 
is to employ frequency-selective geometric control pulses to selectively 
induce desired single-qubit rotations without local addressing.  This 
approach is attractive due to its simplicity, and frequency-selective 
geometric control pulse sequences have been used successfully in nuclear 
magnetic resonance (NMR) applications, and to a lesser extent in electron 
spin resonance experiments.  However, in practice such simple control 
pulse sequences are generally not optimal (for gate operation times and 
other figures of merit), and problems are expected to arise, e.g., when 
the frequency separation between the individual qubits is small, and fast 
control operation is desired to minimize gate operation times and 
deleterious effects of decoherence.  In this case, we expect optimal 
control designs to out-perform the simple geometric schemes.  Indeed, 
simulations show that this is generally the case.  

To consider a concrete example, we compare the Bloch trajectories for 
a simple geometric pulse sequence and an optimally shaped pulse for
the problem of performing simultaneous bit flips on two qubits (here
$1$ and $3$) without affecting the others.  (The bit flip operation
was chosen as it is easy to visualize on the Bloch sphere.  Similar
results hold for other single qubit gates.)  The geometric control
solution in this case is simple: apply a sequence of two $\pi$ pulses 
resonant with the transition frequencies of the two target dots.  If 
the control pulses are sufficiently long, the geometric control pulse 
sequence produces very good results.  However, as we reduce the pulse 
lengths and increase the pulse strengths, off-resonant excitation 
becomes a problem, as Fig.~\ref{fig:QD13Fields} shows.  The left column 
(a) shows the \emph{ideal} evolution of each dot in the rotating frame 
subject to the fields in Fig~\ref{fig:QD13Fields}~(a). Target dots $1$ 
and $3$ follow a smooth path from south to north pole, the others are 
unaffected.  However, the middle column~(b), which shows the actual 
evolution of the quantum dots in the stationary lab frame when subjected 
to the control fields in Fig~\ref{fig:QD13Fields}~(a), shows clearly 
that the applied pulses induce far more complex evolution.  Dots $1$ 
and $3$ are resonantly excited, leading to population transfer from 
the south to the north pole along a spiral path in the stationary frame,
as desired.  However, there is also significant off-resonant excitation 
of the remaining dots, all of which are left in various excited states 
at the final time.  The right column~(c) shows the path of the dots 
subject to field~\ref{fig:QD13Fields}~(b): despite following complicated 
trajectories both target dots finish at the north pole and all the other 
dots are returned to the south pole (ground state) at the final time, 
leaving no unwanted excitations.

\begin{figure}
\scalebox{0.5}{\includegraphics{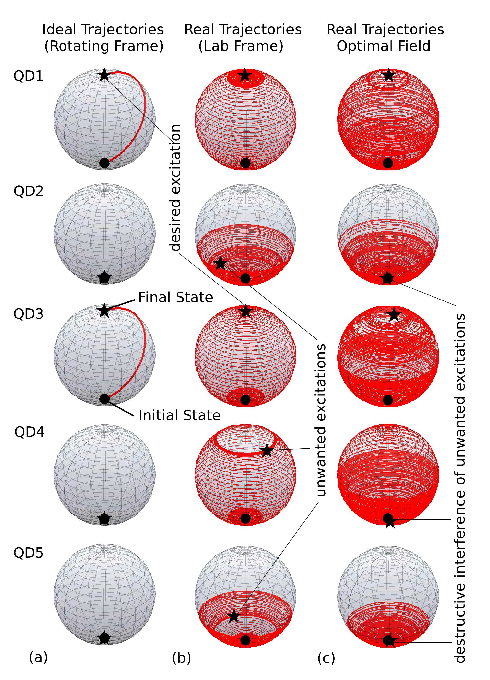}}
\caption{Bloch sphere trajectories for five quantum dots (with resonance
frequencies of $1.32$, $1.35$, $1.375$, $1.38$, $1.397$~eV, respectively)
subject to two Gaussian pi pulses of approximately 2 pico-seconds each
(a,b), and a shaped pulse designed using iterative optimal control (c).
The actual trajectories for the geometric control pulse sequence (b) 
differ significantly from the ideal model trajectories (a).}
\label{fig:QD13Bloch}
\vspace{-2ex}
\end{figure}

The optimal pulse shown in Fig~\ref{fig:QD13Fields}~(b) was designed by
formulating the problem as an optimization problem for the observable
$\op{A}=\diag(0,1,0,-1,0,1,0,-1,0,-1)$, defined on the Hilbert space
$\H=\oplus_{k=1}^5 \H_k$, where $\H_k$ is the single qubit Hilbert space.
Although for five qubits the problem of directly optimizing the average
gate fidelity on the 32-dimensional direct product space $\otimes_{k=1}^5
\H_k$ is computationally still tractable on modern computers, our rather 
unintuitive formulation was considerably more efficient here.  Since we 
assumed inter-dot coupling to be negligible for pico-second pulses in 
our example, it is sufficient to model the evolution of the individual 
dots (for the purpose of implementing single qubit gates) on the direct 
sum Hilbert space $\H=\oplus_{k=1}^5\H_k$, which has only dimension $10$.
Furthermore, the observable $\op{A}$ here assumes its maximum exactly
when dots $1$ and $3$ are in the excited state $\ket{1}$ and all others
are in the ground state $\ket{0}$.  Since all our dots are initially in 
the ground state $\ket{0}$, optimizing $\ave{A}$ requires implementing 
a unitary operator $\op{U}$ such that $\op{U} \ket{0}_k=\ket{1}_k$ for 
$k=1,3$ and $\op{U}\ket{0}_k=\ket{0}_k$ for $k=2,4,5$.  For this simple
example it is easy to see that the only unitary operator (in $\oplus_{k=1}^5
\SU(2)$) that achieves this (on the direct sum Hilbert space) is indeed
$\op{U}=\op{X}\oplus\op{I} \oplus \op{X} \oplus \op{I} \oplus \op{I}$ 
where $X=\begin{pmatrix} 0 & 1\\ 1 & 0 \end{pmatrix}$ is the bit flip 
operator---the operator we wanted to implement.  Similar observables 
can be constructed to implement simultaneous local operations on $N$ 
uncoupled qubits, although the approach does not straightforwardly 
generalize to qudit systems.

The optimal control problem thus defined was solved using the iterative 
optimal control design algorithm described in Sec.~\ref{sec:techniques}.  
The Liouville operator for the system under consideration was
\begin{equation} \label{eq:liouvOp}
  \L[f(t),\rho(t)] = [\op{H}_0 + f(t) \op{H}_1, \op{\rho}(t)],
\end{equation}
where $\op{H}_m=\diag(H_m^{(k)})$, $k=1,\ldots,5$, $m=0,1$, and single 
qubit Hamiltonians 
\[  
  \op{H}_0^{(k)}= \begin{pmatrix} 0 & 0 \\ 0 & \eps_k \end{pmatrix}, \quad 
  \op{H}_1^{(k)}= \frac{1}{\sqrt{2}}\begin{pmatrix} 0 & 1-i \\ 1+i & 0 \end{pmatrix}.
\]  
The optimal pulse shown in Fig.~\ref{fig:QD13Fields}~(b) was obtained 
specifically for $f^{(0)}(t)=0.559\sin(t)$ with $\alpha=\beta=1$ and 
$\lambda=4$.  Convergence for this choice of control parameters was 
comparatively slow (500 iterations), but it was found that choices for 
$\alpha,\beta,\lambda$ that resulted in faster convergence tended to 
produce pulses with undesirable features such as high frequencies or 
high amplitudes, or lower gate fidelities.  The relationship between 
$\alpha,\beta$, $\lambda$ and the initial trial field, and the solution
obtained, in particular its suitability for implementation, versus 
algorithm convergence time are complex, important and worthy of further 
investigation.   

\section{Optimal control: Lyapunov vs iterative control design}
\label{sec:opt:lya}

In the previous section we showed that optimal control design can yield
significant improvements over geometric control for certain problems, and 
that unconventional formulations of a control problem may significantly 
reduce the complexity of a problem and improve the quality of the results.  
In this section, we will compare two basic strategies for optimal control 
design: the iterative approach that was employed in the previous section 
and the mathematically more elegant and computationally less expensive 
approach based on Lyapunov functions.

\begin{figure}
\center\scalebox{0.55}{\includegraphics{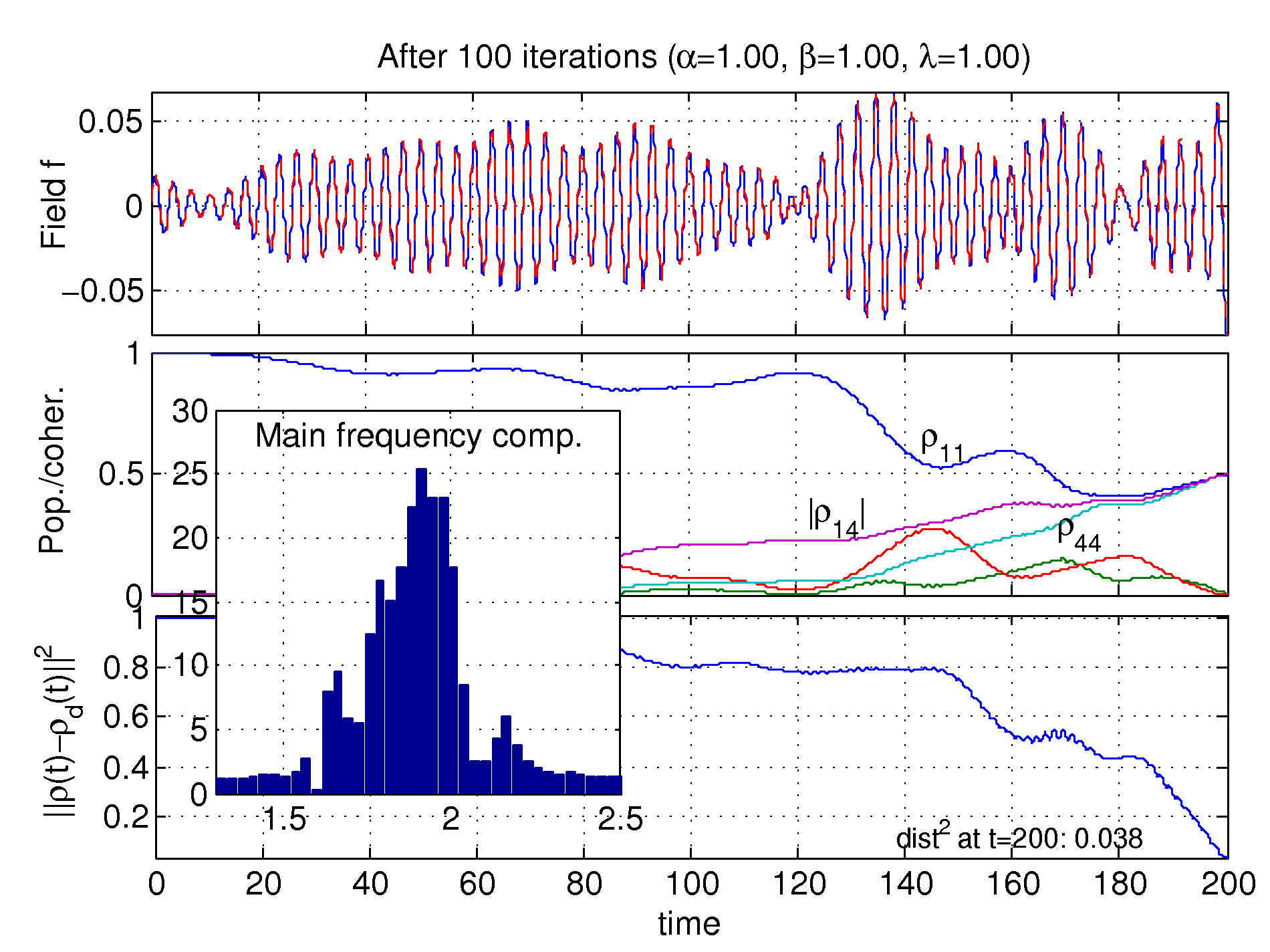}}
\\[2ex]
\center\scalebox{0.55}{\includegraphics{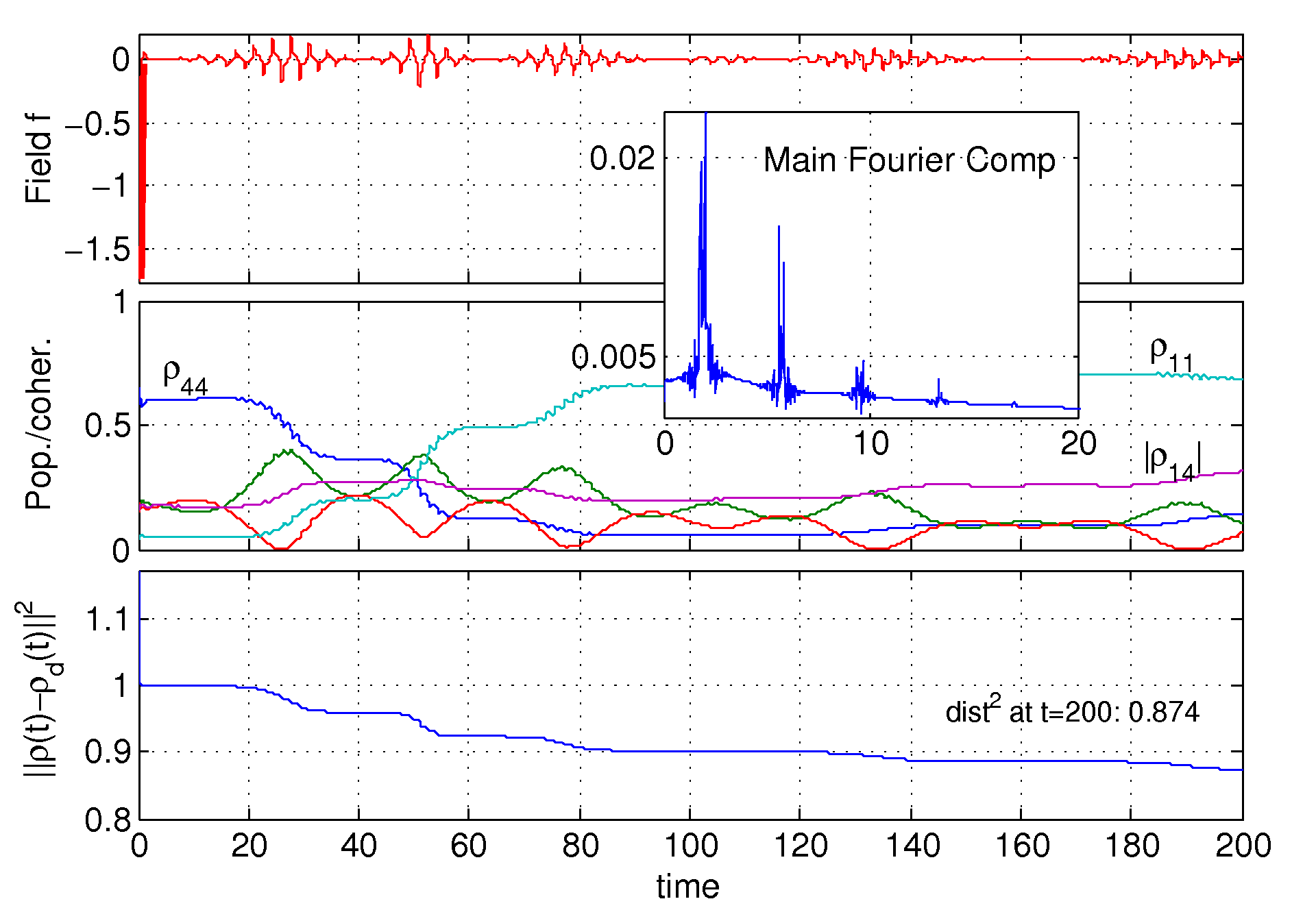}}
\caption{Control pulse, corresponding evolution of the populations $\rho_{kk}$, 
$k=1,2,3,4$, and crucial coherence $|\rho_{14}|$, and distance from the 
target state as a function of time for an optimal control pulse obtained
from an iterative scheme (top) and a Lyapunov-derived pulse (bottom).
Insets shows the main Fourier components of the pulse.}\label{fig:Bell}
\vspace{-1ex}
\end{figure}

Although we could consider the same problem as in the previous section,
for variety we shall instead consider the problem of preparing a Bell 
state $\ket{\Psi_d}=\frac{1}{\sqrt{2}}(\ket{00}+\ket{11})$ starting with 
a product state $\ket{00}$ for two weakly-coupled spins evolving under
the control-dependent Hamiltonian
\begin{equation} \label{eq:sys}
 \op{H}[f(t)] = f(t) \left(0.9 \sigma_x^{(1)}+\sigma_x^{(2)} \right) 
                + 0.1 \sigma_z \otimes \sigma_z,
\end{equation}  
a typical (model) Hamiltonian for systems involving nuclear spins, e.g.,
in NMR.  In this case we formulated the problem as a straightforward 
state control problem.  Fig.~\ref{fig:Bell} (top) shows the results
produced by the iterative optimal control algorithm for $\op{A}$ chosen 
to be the projection onto the target state and $\alpha=\beta=\lambda=1$ 
for a target pulse duration of $t_F=200$ time units.  While the correct 
choice of the parameters $\alpha$, $\beta$, $\lambda,f^{(0)}$, \ldots in 
the iterative scheme appears to be crucial to obtain good results, and 
finding suitable parameters is not always a trivial task, in most cases 
it was obvious very quickly when we had chosen the parameters poorly,
allowing us to adjust the parameters.  After some tuning, the control
pulses obtained from the iterative scheme were efficient in steering the
system very close to the target state and had desirable characteristics,
as the example shown in Fig.~\ref{fig:Bell} shows, despite the fact that
system~(\ref{eq:sys}) is not controllable.  The distance from the target 
state at the specified target time is very small, and the plot of the 
populations and relevant coherence $|\rho_{14}|$ shows smooth
trajectories approaching the target values almost monotonically.  Thus,
the pulse is highly efficient in steering the system to the target state, 
near energy optimal, and has desirable characteristics such as limited
variation in amplitude (no spikes) and a narrow frequency spectrum 
centered around the transition frequencies of the system.

The Lyapunov potential approach proved considerably more problematic.  
Firstly, many initial states of interest such as the separable state
$\ket{00}$ in our example, lie within the set of states for which 
convergence does not occur as $\bra{\Psi_d(0)}H_1\ket{\Psi(0)}=0$, and 
hence the Lyapunov derived control field $f\equiv 0$.  We can overcome 
this problem by applying a random initial pulse to kick the system out 
of its initial equilibrium state.  Even with this modification, however, 
in all our experiments the controls derived from the Lyapunov potential 
function performed significantly worse than those derived from iterative 
schemes, both in terms of the rate of convergence to the target state, 
and the characteristics of the pulses.  In fact, in our example, the 
distance from the target trajectory, although monotonically decreasing 
as a function of time, exhibits plateaux where the system appears to be 
trapped in 'false' semi-stable states for long periods of time.  
Furthermore, the control fields we obtained, a typical example of which 
is shown in Fig.~\ref{fig:Bell} (bottom), tended to be generally spiky 
and spectrally more complex than those derived from iterative schemes.

In our particular example (and a number of other cases we have studied 
so far) we found that the iterative algorithm allowed us to find control 
pulses that steer the system to a state very close to the target state
in a relatively short time, as Fig.~\ref{fig:Bell} shows.  This applied
even to some systems that are not generally controllable, provided that
the control objective was dynamically realizable, and in cases where it
was not attainable, we were often able to steer the system to reachable
states that maximized the expectation value of the target observable
subject to the dynamical constaints.

The local optimization approach on the other hand, despite recent claims
of strong global convergence \cite{qph0601016,qph0506268}, seemed to fare 
significantly worse in our simulations.  The system often got trapped for 
long periods in semi-stable critical points, and convergence to the target 
state was slow at best.  Thus, while asymptotic convergence may be an 
attractive property mathematically, it appears to be less useful in 
practice, especially where gate-operation and state preparation times are 
critical, although further work is necessary to assess whether the cases
we have studied are anomalous, and whether the performance of the local 
gradient optimization can be improved, e.g., by choosing different Lyapunov 
potentials.  The latter question is significant as local optimization
techniques can in principle be adapted for measurement-based based feedback
control in the context of stochastic differential
equations~\cite{qph0510222, SIAMJCO46p445}.

\section{Conclusion}

We have considered different ways of modelling quantum control systems,
formulating open-loop control problems and techniques for solving them.
The preliminary results presented here strongly suggest that model-based 
optimal control design tends to produce control pulses that are more 
effective, robust and efficient, than simpler schemes based on geometric 
pulses.  However, the quality of the pulses obtained can vary significantly 
depending on the choice of model, problem formulation and the technique 
used for control design.  

We have seen in particular that control design based on instantaneous 
minimization of a Lyapunov potential function, despite attractive features 
such as simplicity and asymptotic convergence properties, tends to produce fields inferior to those obtained from common iterative optimal control 
schemes in various regards, from rate of convergence to robustness and 
what might be termed energy and spectral optimality.  Preliminary results
suggest that iterative optimal control design produces far better results.

Nonetheless, instantaneous optimization techniques remain of interest due 
to the potential to adapt these techniques in the context of model-based
feedback control.  More work is necessary to understand how model choice, 
problem formulation and algorithmic parameters affect the convergence 
behavior and characteristics of the fields obtained, in particular for 
instantaneous optimization techniques, and how these characteristics might 
be improved, e.g., by choosing more sophisticated Lyapunov potentials.


\begin{thebibliography}{15}
\expandafter\ifx\csname natexlab\endcsname\relax\def\natexlab#1{#1}\fi
\expandafter\ifx\csname bibnamefont\endcsname\relax
  \def\bibnamefont#1{#1}\fi
\expandafter\ifx\csname bibfnamefont\endcsname\relax
  \def\bibfnamefont#1{#1}\fi
\expandafter\ifx\csname citenamefont\endcsname\relax
  \def\citenamefont#1{#1}\fi
\expandafter\ifx\csname url\endcsname\relax
  \def\url#1{\texttt{#1}}\fi
\expandafter\ifx\csname urlprefix\endcsname\relax\def\urlprefix{URL }\fi
\providecommand{\bibinfo}[2]{#2}
\providecommand{\eprint}[2][]{\url{#2}}

\bibitem[1]{PRA49p2133}
\bibinfo{author}{\bibfnamefont{H.~M.} \bibnamefont{Wiseman}},
\emph{Quantum theory of continuous feedback},
  \bibinfo{journal}{Phys. Rev. A} \textbf{\bibinfo{volume}{49}},
  \bibinfo{pages}{2133} (\bibinfo{year}{1994}).

\bibitem[2]{NAT417p533}
\bibinfo{author}{\bibfnamefont{J.~L.}~\bibnamefont{Herek \emph{et al.}}},
\emph{Quantum control of energy flow in light.}
  \bibinfo{journal}{Nature} \textbf{\bibinfo{volume}{417}},
  \bibinfo{pages}{533} (\bibinfo{year}{2002}).

\bibitem[3]{SCI282p919}
\bibinfo{author}{\bibfnamefont{A.}~\bibnamefont{Assion} \emph{et al.}},
\emph{Control of chemical reactions by feedback-optimized phase-shaped
	   femtosecond laser pulses},
  \bibinfo{journal}{Science} \textbf{\bibinfo{volume}{282}},
  \bibinfo{pages}{919} (\bibinfo{year}{1998}).

\bibitem[4]{PRE70n016704}
\bibinfo{author}{\bibfnamefont{G.}~\bibnamefont{Turinici}},
  \bibinfo{author}{\bibfnamefont{C.}~\bibnamefont{{Le Bris}}},
  \bibinfo{author}{\bibfnamefont{H.}~\bibnamefont{Rabitz}},
\emph{Efficient algorithms for laboratory discovery of optimal control controls},
  \bibinfo{journal}{Phys. Rev. E} \textbf{\bibinfo{volume}{70}},
  \bibinfo{pages}{016704} (\bibinfo{year}{2004}).

\bibitem[5]{PRA63n063412}
\bibinfo{author}{\bibfnamefont{B.~J.} \bibnamefont{Pearson}},
  \bibinfo{author}{\bibfnamefont{J.~L.} \bibnamefont{White}},
  \bibinfo{author}{\bibfnamefont{T.~C.} \bibnamefont{Weinacht}},
  \bibinfo{author}{\bibfnamefont{P.~H.} \bibnamefont{Bucksbaum}}, 
\emph{Coherent control using adaptive learning algorithms},
\bibinfo{journal}{Phys. Rev. A}
  \textbf{\bibinfo{volume}{63}}, \bibinfo{pages}{063412}
  (\bibinfo{year}{2001}).

\bibitem[6]{PRA63n032308}
\bibinfo{author}{\bibfnamefont{N.}~\bibnamefont{Khaneja}},
  \bibinfo{author}{\bibfnamefont{R.}~\bibnamefont{Brockett}}, 
  \bibinfo{author}{\bibfnamefont{S.~J.} \bibnamefont{Glaser}},
\emph{Time-optimal control in spin systems},
  \bibinfo{journal}{Phys. Rev. A} \textbf{\bibinfo{volume}{63}},
  \bibinfo{pages}{032308} (\bibinfo{year}{2001}).

\bibitem[7]{JCP110p9825}
\bibinfo{author}{\bibfnamefont{Y.}~\bibnamefont{Ohtsuki}},
  \bibinfo{author}{\bibfnamefont{W.}~\bibnamefont{Zhu}}, 
  \bibinfo{author}{\bibfnamefont{H.}~\bibnamefont{Rabitz}},
\emph{Monotonically convergent algorithm for quantum control with dissipation},
  \bibinfo{journal}{J. Chem. Phys.} \textbf{\bibinfo{volume}{110}},
  \bibinfo{pages}{9825} (\bibinfo{year}{1999}).

\bibitem[8]{PRA61n012101}
\bibinfo{author}{\bibfnamefont{S.~G.} \bibnamefont{Schirmer}},
  \bibinfo{author}{\bibfnamefont{M.~D.} \bibnamefont{Girardeau}},
  \bibinfo{author}{\bibfnamefont{J.~V.}~\bibnamefont{Leahy}},
\emph{Efficient algorithm for optimal control of mixed-state systems},
  \bibinfo{journal}{Phys. Rev. A} \textbf{\bibinfo{volume}{61}},
  \bibinfo{pages}{012101} (\bibinfo{year}{2000}).

\bibitem[9]{JCP118p8191}
\bibinfo{author}{\bibfnamefont{Y.}~\bibnamefont{Maday}},
  \bibinfo{author}{\bibfnamefont{G.}~\bibnamefont{Turinici}},
\emph{New formulations of monotonically convergent quantum control algorithms},
  \bibinfo{journal}{J. Chem. Phys.} \textbf{\bibinfo{volume}{118}},
  \bibinfo{pages}{8191} (\bibinfo{year}{2003}).

\bibitem[10]{JMR172p296}
\bibinfo{author}{\bibfnamefont{N.}~\bibnamefont{Khaneja}},
  \bibinfo{author}{\bibfnamefont{T.}~\bibnamefont{Reiss}},
  \bibinfo{author}{\bibfnamefont{C.}~\bibnamefont{Kehlet}},
  \bibinfo{author}{\bibfnamefont{T.}~\bibnamefont{Schulte-Herbrueggen}},
  \bibinfo{author}{\bibfnamefont{S.~J.}~\bibnamefont{Glaser}}, 
\emph{Optimal control of coupled spin dynamics: design of NMR pulse
	   sequences by gradient ascent algorithms},
\bibinfo{journal}{J. Mag. Resonance}
  \textbf{\bibinfo{volume}{172}}, \bibinfo{pages}{296} (\bibinfo{year}{2005}).

\bibitem[11]{Mirrahimi04}
\bibinfo{author}{\bibfnamefont{M.}~\bibnamefont{Mirrahimi}},
   \bibinfo{author}{\bibfnamefont{P.}~\bibnamefont{Rouchon}}, 
\emph{Trajectory tracking for quantum systems: a Lyapunov approach},
in \emph{\bibinfo{booktitle}{Proceedings of the {MTNS}04}}
  (\bibinfo{year}{2004}).

\bibitem[12]{qph0601016}
\bibinfo{author}{\bibfnamefont{C.}~\bibnamefont{Altafini}},
  \emph{\bibinfo{title}{Feedback control of spin systems}},
  quant-ph/0601016  (\bibinfo{year}{2006}).

\bibitem[13]{qph0506268}
\bibinfo{author}{\bibfnamefont{C.}~\bibnamefont{Altafini}},
\emph{\bibinfo{title}{Feedback stabilization of quantum ensembles: a
  global convergence analysis on manifolds}},
  quant-ph/0506268 (\bibinfo{year}{2005}).

\bibitem[14]{qph0510222}
\bibinfo{author}{\bibfnamefont{C.}~\bibnamefont{Altafini}},
\bibinfo{author}{\bibfnamefont{F.}~\bibnamefont{Ticozzi}},
\emph{\bibinfo{title}{Almost global stochastic feedback stabilization of
  conditional quantum dynamics}},
  quant-ph/0510222 (\bibinfo{year}{2005}).

\bibitem[15]{SIAMJCO46p445}
\bibinfo{author}{\bibfnamefont{M.}~\bibnamefont{Mirrahimi}},
\bibinfo{author}{\bibfnamefont{R.}~\bibnamefont{von Handel}},
\emph{\bibinfo{title}{Stabilizing feedback controls for quantum systems}},
\bibinfo{journal}{SIAM J. Control Optim.}
\textbf{\bibinfo{volume}{46}}, \bibinfo{pages}{445-467}
(\bibinfo{year}{2007}).

\end{thebibliography}

\end{document}